\documentclass[final]{svjour3}
\usepackage{graphicx}
\usepackage{rotating}
\usepackage{amsmath, amssymb}
\usepackage{upgreek}
\usepackage{bm}
\usepackage{geometry}
\usepackage{float}
\usepackage[numbers]{natbib}
\makeatletter
\journalname{Journal of Low Temperature Physics}
\bibpunct{}{}{,}{s}{}{,}

\def\lb{\emph{LiteBIRD}}

\begin{document}

\title{Monitoring TES Loop Gain in Frequency Multiplexed Readout}

\titlerunning{TES Loop Gain Monitor in fMux}

\author{T.~de~Haan$^{1,2}$ \and 
        T.~Adkins$^{3}$ \and
        M.~Hazumi$^{1,2}$ \and
        D.~Kaneko$^{2}$ \and
        J.~Montgomery$^{4}$ \and
        G.~Smecher$^{4,5}$ \and
        A.~Suzuki$^{6}$ \and
        Y.~Zhou$^{2}$}

\authorrunning{T.~de~Haan et al.}

\institute{$^1$ Institute of Particle and Nuclear Studies (IPNS), High Energy Accelerator Research Organization (KEK), Tsukuba, Ibaraki, 305-0801, Japan \\
           $^2$ International Center for Quantum-field Measurement Systems for Studies of the Universe and Particles (QUP-WPI), High Energy Accelerator Research Organization (KEK), Tsukuba, Ibaraki, 305-0801, Japan \\
           $^3$ Department of Physics, University of California, Berkeley, CA, 94720, USA \\
           $^4$ Department of Physics and McGill Space Institute, McGill University, 3600 Rue University, Montreal,
QC H3A 2T8, Canada \\
           $^5$ Three-Speed Logic, Inc., Victoria, BC, V8S 3Z5 Canada \\
           $^6$ Lawrence Berkeley National Laboratory (LBNL), Berkeley, CA, 94720, USA \\
\email{tijmen@post.kek.jp}}

\maketitle

\begin{abstract}
    We present a method for precise monitoring of the loop gain of transition edge sensors (TES) under electrothermal feedback. The measurement is implemented on the ICE DfMux electronics and operates simultaneously with Digital Active Nulling (DAN). It uses one additional bias sinusoid per TES and does not require any additional readout channels. The loop gain monitor is being implemented on the Simons Array and is an integral part of the baseline calibration strategy for the upcoming \lb\ satellite.
\end{abstract}

\section{Motivation}

Precise calibration from readout counts to power units over a wide range of timescales poses a major challenge for the operation of transition edge sensors (TES). The calibrations can vary over time due to environmental factors such as changes in optical loading, magnetic field, or thermal bath temperature. For ground-based cosmic microwave background (CMB) telescopes, various solutions have been used to monitor the calibration over time. Arguably the most successful solution has been to periodically inject chopped optical power during special calibration observation periods that typically last a few minutes. While this technique does not provide a consistent signal from detector-to-detector, or an absolute calibration, the injected power is fairly stable in time and extremely useful in monitoring time variation of per-detector calibrations.

The CMB experiment with one of the most stringent calibration stability requirements is \lb, a space-based CMB observatory that will measure the tensor-to-scalar ratio $r$ with a total uncertainty of $\delta r = 1 \times 10^{-3}$, which includes statistical uncertainty, systematic uncertainty, and margin. The control of systematic uncertainties is extremely important, and the requirements on gain fluctuations are accordingly stringent\cite{ptep22}. Unfortunately, \lb\ will not include a chopped optical source, so there is a strong need for an alternative way to monitor calibration over time.

In this work, we present a method for monitoring the readout count-to-power calibration by precisely monitoring the loop gain of TES electrothermal feedback through injection of one additional sideband for each channel. The method is explained in \S\ref{sec:theory}. In \S\ref{sec:implementation}, we describe our practical implementation and show that it works very well. The always-on version of the loop gain measurement, termed the loop gain monitor, is presented in \S\ref{sec:lgm}.

\section{Loop Gain Measurement}
\label{sec:theory}

Digital frequency multiplexing of voltage-biased TES bolometers (DfMux) uses 6 wires to control a SQUID amplifier in series with a bank of LC resonators and TES detectors. One pair of wires is connected to the SQUID output, one is connected to the SQUID input, and the final pair is used to drive a sum of sinusoids which ultimately bias the detector. Each sinusoid is precisely tuned in frequency to one LC resonant frequency as to address exactly one detector.

In the method we propose, each detector channel---regardless of whether it reads out an optical TES, dark TES, or a calibration resistor---is driven by a pair of sinusoids. The primary sinusoid, known as the ``carrier,'' provides the bias power necessary to maintain the detector at its desired operating point. Accompanying this, a secondary sinusoid, with significantly lower amplitude and offset by a frequency $\delta \omega$ from the carrier (typically corresponding to a few Hz), is also introduced. As a result, the power deposited on the TES is predominantly steady-state, with a small modulation at the angular ``beat frequency'' of $\delta \omega$.

Let's now calculate how we obtain the detector loop gain $\mathcal{L}$ from the measured current. The starting point will be the power balance equation. We assume that the detector is biased with an ideal voltage bias $V$, though we note that the final result holds for any Th\'evenin equivalent circuit. We will also assume throughout that we inject an upper sideband voltage $\delta V$ with $\delta V \ll V$. Furthermore assuming we choose a value of $\delta \omega$ that corresponds to a timescale much slower than the time constant of the TES, we can assume equilibrium and write
\begin{equation}
    \langle P_\mathrm{opt} \rangle + \frac{\langle \left( V \cos{(\omega t)} + \delta V \cos{(\left(\omega + \delta \omega \right) t)} \right)^2 \rangle}{R(t)} = K \left( T(t)^n - T_\mathrm{bath}^n \right)
    \label{eq:power_balance}
\end{equation}
where $P_\mathrm{opt}$ is the optical power incident on the TES, and $K$ and $n$ provide the usual parameterization for thermal conductance \cite{irwin95}. The angled brackets denote averaging over the aforementioned timescale. $R(t)$ and $T(t)$ are the dynamic variables that are related by the intrinsic R--T relationship of the TES. 

Let's expand the relevant variables to first order
\begin{equation}
    \begin{aligned}
        V(t) = V \cos{(\omega t)} + \delta V \cos{((\omega + \delta \omega) t)} \\
        I(t) = I \cos{(\omega t)} + \delta I_+ \cos{((\omega + \delta \omega) t)} + \delta I_- \cos{((\omega - \delta \omega) t)} \\
        R(t) = R + \delta R \cos{(\delta \omega t)} \\
        T(t) = T + \delta T \cos{(\delta \omega t)}
    \end{aligned}
\end{equation}

As has been done many times in the literature\cite{lee98}, it can be shown that these assumptions give rise to an analogy to the open loop gain of an amplifier under negative feedback. Here, we find that the loop gain is given by
\begin{equation}
  \mathcal{L}= \frac{\alpha P_\mathrm{electrical}}{K n T^n} = \frac{\alpha \frac{1}{2} V^2}{K n T^n R} \ \ .
  \label{eq:loop_gain}
\end{equation}
where $\alpha$ is defined as the logarithmic derivative of $R$ with respect to $T$ i.e. $\alpha = T \delta R / R \delta T$. Note that this is only valid in the case of ideal voltage bias.
    
\begin{figure}
\begin{center}
\includegraphics[width=\linewidth,keepaspectratio]{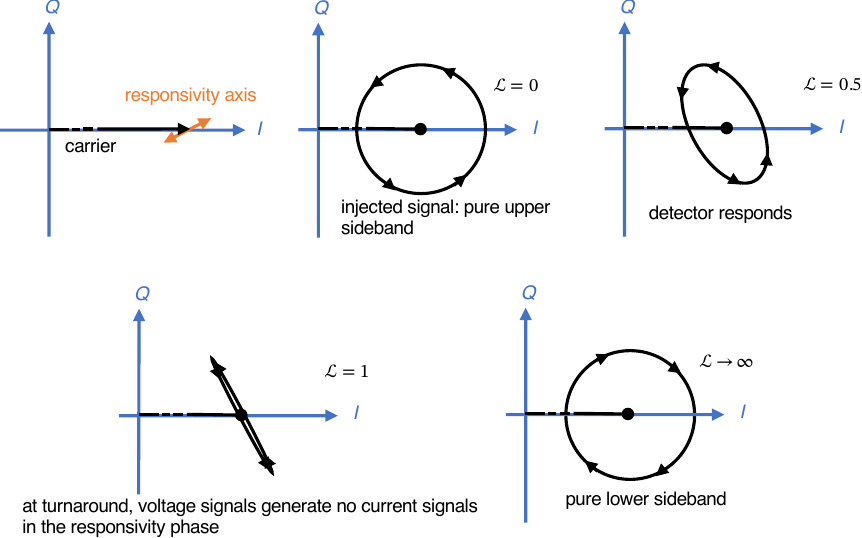}
\end{center}
\caption{Diagram illustrating the loop gain measurement process. Initially, the carrier voltage takes a large positive value of the in-phase, or ``I'' current, by the definition of the I and Q phases. A minor voltage sideband, just above the carrier frequency, is introduced. In complex-demodulated space, this ``upper sideband'' manifests as counterclockwise motion around the bias point. As the TES starts to exhibit increasingly strong electrothermal feedback, the motion changes according to the loop gain of the feedback. In the limit of high loop gain, the motion becomes a clockwise circle, signifying a pure lower sideband.}
\label{fig:explanation}
\end{figure}

Taylor expanding Equation \ref{eq:power_balance} to first order and making use of the expression for loop gain from Equation \ref{eq:loop_gain} we find that
\begin{equation*}
    \delta R = 2 R \frac{\delta V}{V} \frac{\mathcal{L}}{1+\mathcal{L}}
\end{equation*}
The measured quantity is the current $I(t)$, which we can calculate using Ohm's law $I(t) = V(t)/R(t)$. Again considering only the first-order terms of the Taylor expansion yields
\begin{equation}
    \begin{aligned}
        \delta I_- = - \frac{\delta V}{R} \frac{\mathcal{L}}{1+\mathcal{L}} \\
        \delta I_+ = \frac{\delta V}{R} \frac{1}{1+\mathcal{L}}
    \end{aligned}
    \label{eq:loop_gain_measurement}
\end{equation}
where we observe that the resistance modulation of the carrier influences both the upper and lower sidebands. An additional term due to the injected voltage sideband also contributes to the upper sideband. Thus, in the absence of electrothermal feedback ($\mathcal{L} \rightarrow 0$), the only detected signal is the initial upper sideband. However, as the TES enters the transition, the upper sideband becomes suppressed, giving rise to a lower sideband. When loop gain is high ($\mathcal{L} \rightarrow \infty$), the upper sideband is completely suppressed, with the lower sideband reaching the same amplitude as the initially detected upper sideband. See Figure \ref{fig:explanation} for a visualization of this effect in complex-demodulated space.

By taking the \textit{ratio} of the lower to the upper sideband, equivalent to $|\delta I_-|/|\delta I_+| = \mathcal{L}$, we achieve our loop gain measurement. By its nature as a ratio measurement, it is inherently robust against drifts in calibration, as any such drifts would affect $|\delta I_-|$ and $|\delta I_+|$ equally.

\section{Experimental Implementation}
\label{sec:implementation}

We have implemented the loop gain measurement on the DfMux version of the ICE readout electronics \cite{bandura16}. The primary complication was the presence of Digital Active Nulling (DAN) \cite{dehaan12}, which means that we cannot use a secondary listening channel to extract the upper and lower sideband signals separately, as they are largely nulled away. This makes previous-generation homodyne implementations of sideband measurements impossible, instead requiring a full heterodyne understanding of sidebands using the DAN signals directly. DAN returns time-ordered data for both the in-phase and quadrature demodulation. We cast the measured current as a complex data stream $d(t) = I_\mathrm{DAN, in-phase}(t) + i I_\mathrm{DAN, quadrature}(t)$, and extract the upper and lower sidebands as follows:
\begin{equation}
    \begin{aligned}
    \delta I_- = \langle \left( d(t) - \langle d(t) \rangle \right) \exp{i \delta \omega t} \rangle \\
    \delta I_+ = \langle \left( d(t) - \langle d(t) \rangle \right) \exp{-i \delta \omega t} \rangle
    \end{aligned}
    \label{eq:demodulation}
\end{equation}
and calculate the loop gain as $\mathcal{L} = \left| \delta I_- / \delta I_+ \right|$. It should be noted that taking the absolute value of the measured loop gain is disregarding potentially valuable phase information. This phase information and its dependence on $\delta \omega$ is the topic of a companion paper \cite{zhou23}.

\subsection{Loop Gain Measurement}
\label{sec:measurement}

Our experiment features a CIMM (Cold Integrated fMux Module) \cite{dehaan19} with a $40\times$ LC chip based on the design by Rotermund et al. \cite{rotermund16}, but fabricated using niobium instead of aluminum to reduce the effective series resistance (ESR), 112-junction SSAA (Series SQUID Array Amplifier) from STAR Cryoelectronics \cite{boyd17}, ten $1 \Upomega$ calibration resistors, and 30 TES bolometers fabricated by SeeQC \cite{suzuki20}.

\begin{figure}
\begin{center}
\includegraphics[width=0.8\linewidth,keepaspectratio, trim={0cm 0.2cm 0cm 0cm},clip]{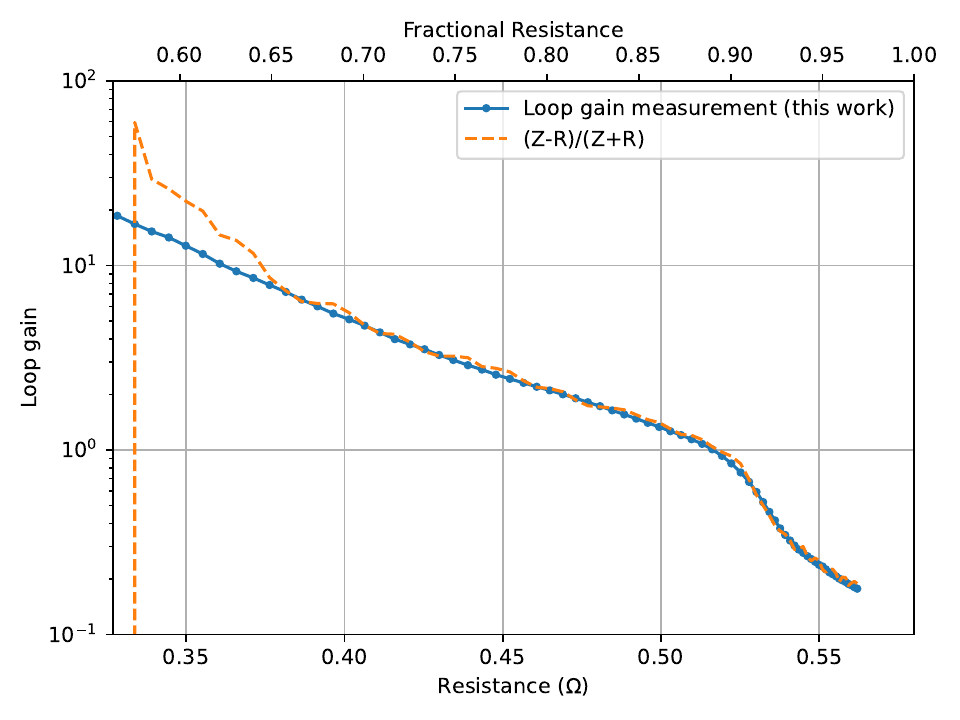}
\end{center}
\caption{Loop gain measurements for one TES shown in blue. The orange dashed curve shows the loop gain that one might estimate directly from the IV curve, although this estimate fails deep in the transition due to the assumption of ideal voltage bias.}
\label{fig:measurement}
\end{figure}

Figure \ref{fig:measurement} shows a representative loop gain measurement from one of the TES channels. We find the measurement to be reliably replicable, leading us to have confidence in its accuracy. For comparison, we also show the loop gain estimated from the IV curve assuming perfect voltage bias: $(Z-R)/(Z+R)$, where $Z$ and $R$ refer to the dynamic impedance $dV/d|I|$ and total impedance $V/|I|$, respectively \cite{elleflot20}. While this method agrees with our loop gain measurement higher up in the transition, it fails as the detector drops deeper into its transition by diverging and eventually giving negative values. This effect can be attributed to the fact that the denominator $Z+R$ diverges as $VI \rightarrow \mathrm{constant}$, and can go negative if there is an additional voltage drop due to any a series parasitic impedance. In contrast, our loop gain measurement does not suffer from this effect as it empirically probes the effect of electrothermal feedback and does not rely on any knowledge of series parasitic impedances. 

\section{Loop Gain Monitor}
\label{sec:lgm}

\begin{figure}
\begin{center}
\includegraphics[width=\linewidth,keepaspectratio]{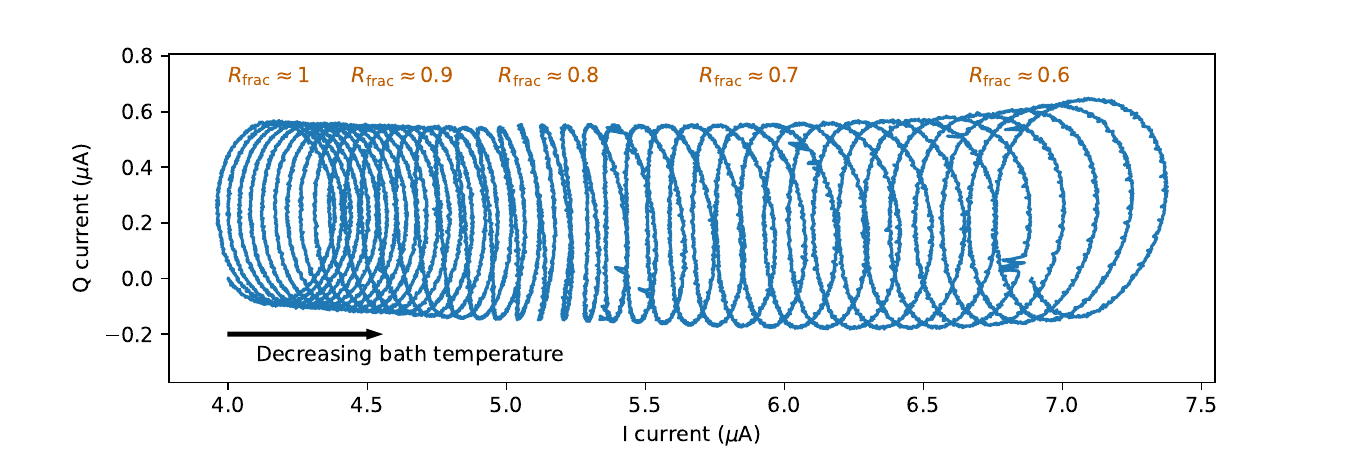}
\end{center}
\caption{Demonstration of the loop gain monitor. The bath temperature was varied while the voltage bias was held at $2.8~\mu \mathrm{V}$. Starting at the left, when the resistance is high and therefore the in-phase ($I$) current is relatively low, the loop gain monitor shows counterclockwise motion, indicating $\mathcal{L}<1$. As the TES drops into transition, it crosses $\mathcal{L}=1$, where there is no motion along the responsivity axis. Finally, as the TES drops deeper into transition, the motion becomes clockwise, indicating $\mathcal{L}>1$.}
\label{fig:lgm1}
\end{figure}

We introduce the loop gain monitor (LGM), which is an always-on version of the loop gain measurement described in \S\ref{sec:measurement}. The LGM is designed to operate continuously during regular data taking. Figure~\ref{fig:lgm1} shows an illustrative demonstration of the LGM, where bath temperature is changed rapidly and drastically in order to show how the LGM operates.

We reduce the time-ordered data as follows. We split the time-ordered data into chunks during which exactly one cycle of the LGM has occured. For this time chunk, we define a Vandermonde-like matrix with rows containing the known components $\bm{V}_i = \left[ \exp{-2 \pi i \delta \omega t_i}, \exp{2 \pi i \delta \omega t_i}, 1 \right]$. We then obtain the best-fit parameters by calculating the linear least squares solution

\begin{equation}
\hat{\bm{p}} = (\bm{V}^\top \bm{V})^{-1} \bm{V}^\top \bm{d}
\end{equation}

where $\bm{d}$ is the vector of time-ordered data corresponding to the relevant time chunk of $d(t)$ from Eq.~\ref{eq:demodulation}. Once we have obtained these best-fit parameter estimates $\hat{\bm{p}}$, we take the ratio of its first and second elements to be the loop gain estimate. By repeating this for every time chunk, we obtain an estimate of loop gain over time which, in turn, can be used to predict the evolution of the TES calibration over time. Figure \ref{fig:lgm2} shows this prediction for data taken in the laboratory. It is unknown how much of the variation in this measurement is due to noise and how much due to true variations in loop gain. However, we can place an upper limit on the measurement noise by making the conservative assumption that all measured variation is due to measurement noise. We then calculate the amplitude spectral density of the loop gain data under the white noise assumption. We assume Gaussian statistics to scale between different integration times and convert to a responsivity $S\propto\mathcal{L}/(\mathcal{L}+1)$. Consequently, we scale the loop gain uncertainty from the laboratory configuration to the expected flight performance using the ratio of the measured laboratory current noise ($120~\mathrm{pA}/\sqrt{\mathrm{Hz}}$) to the expected \lb\ current noise ($8~\mathrm{pA}/\sqrt{\mathrm{Hz}}$):
\begin{equation}
\mathrm{ASD}(S)_\textit{LiteBIRD} \approx \mathrm{ASD}(S) \times \frac{8~\mathrm{pA}/\sqrt{\mathrm{Hz}}}{120~\mathrm{pA}/\sqrt{\mathrm{Hz}}} = 8 \times 10^{-3} /\sqrt{\mathrm{Hz}}  .
\end{equation}
This corresponds to a precision of $3\times10^{-5}$ each 24-hour period, which exceeds even the most strict \lb{} calibration requirement\cite{ghigna20} of $10^{-4}$.

{The \lb\ collaboration calculated an allowable level of 1/f multiplicative noise---also known as gain fluctuation\cite{ptep22} in terms of a focal plane temperature fluctuation power spectrum. We perform a numerical simulation based on the TES power balance equation and cast this allowable level in terms of TES responsivity fluctuations, yielding a power spectrum of $5\times10^{-5} (1 \mathrm{Hz}/f)$. We compare the lower bound on LGM performance to this level of allowable 1/f multiplicative noise at the CMB temperature dipole frequency of the inverse scan rate (1/20 minutes), where the loop gain monitor will be compared to the CMB dipole amplitude. We find that our conservative measure of the LGM performance falls within the acceptable bound for the \lb\ mission objectives. We expect that a more sophisticated analysis of the LGM data will further improve the uncertainties, granting additional margin.

\begin{figure}
\begin{center}
\includegraphics[width=\linewidth,keepaspectratio]{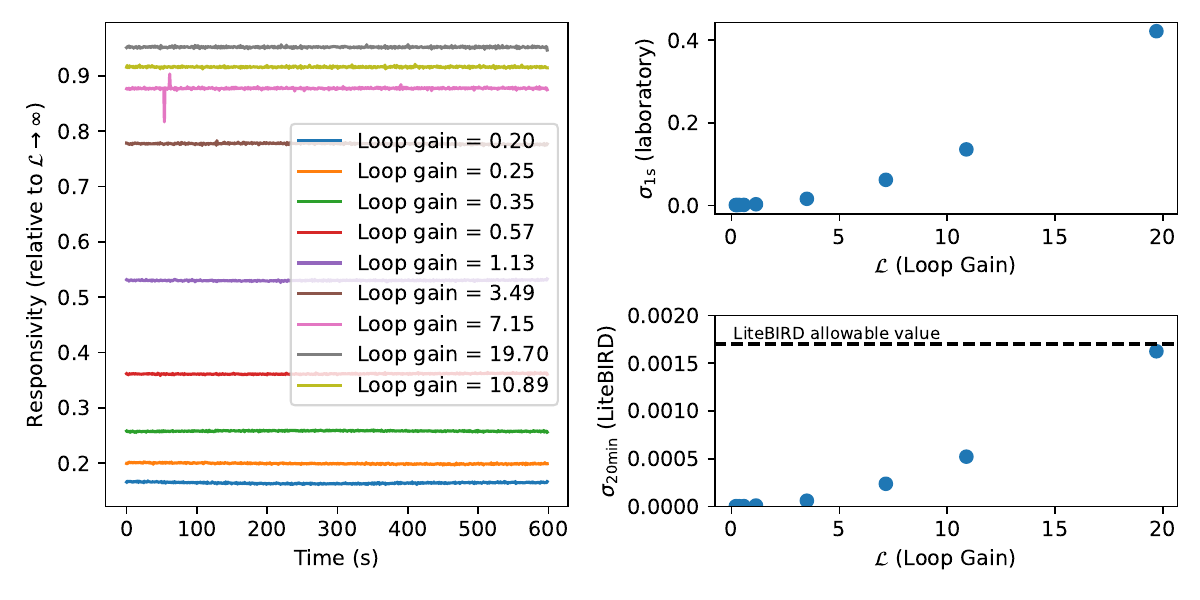}
\end{center}
\caption{Loop gain monitor-derived relative calibration over time, as measured in the laboratory. Assuming all variation is due to measurement noise, and that the noise scales according to Gaussian statistics, we derive a calibration precision of $8\times10^{-3}/\sqrt{\mathrm{Hz}}$ assuming \lb\ noise levels. On 20-minute timescales, where the \lb\ calibration is anchored to the CMB dipole, the extrapolated LGM performance is sufficient for a conservative approximation of the \lb\ allowable value.}
\label{fig:lgm2}
\end{figure}

For an experiment where the full TES bandwidth is used for science, such as in TES X-ray microcalorimeters, there will be a small amount of contamination due to the LGM. Since the LGM operates at a very well defined frequency, we expect that a simple notch filter can be used to remove such contamination with minimal impact on scientific objectives. In the case of, e.g., \lb, the combination of a large point-spread function and relatively slow scanning means that the science bandwidth is very limited. In fact, the maximum required frequency for science data is only a few Hz, set by the rotation rate of the half-wave plate polarization modulator. As such, the LGM can be used without any overlap with the CMB signal at all.

\section{Conclusions and Outlook}
\label{sec:conclusions}

In this work, we have presented a precise and novel method for monitoring the loop gain of a TES under electrothermal feedback. Our method is operationally compatible with Digital Active Nulling, and only requires a single additional bias sinusoid per TES. With its implementation on the ICE DfMux electronics, the loop gain monitor has proven to be a successful solution to the challenges posed by calibration stability requirements for experiments such as \lb.

The LGM, which ultimately consists of a ratio of measurements, is, therefore, inherently robust against any drifts in calibration. We find that the method provides accurate results that are reliably replicable over a range of conditions. The statistical uncertainty of the LGM is excellent and exceeds any reasonable requirements. 

The practical implementation of the LGM is underway for the Simons Array, where the intention is to demonstrate that the LGM-derived calibrations match the calibrations measured using the optical chopped source. This will disambiguate true loop gain variation from uncertainty in the loop gain monitor. \lb\ has adopted the LGM as its baseline strategy for monitoring calibrations over time, relying on the CMB dipole to measure the absolute calibration no more frequently than once per 6 months. 

\begin{acknowledgements}
This work was supported by World Premier International Research Center Initiative (WPI), MEXT, Japan.
\end{acknowledgements}

\pagebreak

\end{document}